\documentclass{vgtc}                          

\ifpdf
  \pdfoutput=1\relax                   
  \usepackage{graphicx}                
  \DeclareGraphicsExtensions{.pdf,.png,.jpg,.jpeg} 
\else
  \usepackage{graphicx}                
  \DeclareGraphicsExtensions{.eps}     
\fi%

\graphicspath{{figures/}{pictures/}{images/}{./}} 

\usepackage{microtype}
\PassOptionsToPackage{warn}{textcomp}
\usepackage{textcomp}
\DeclareMathAlphabet{\altmathcal}{OMS}{cmsy}{m}{n}
\usepackage{mathptmx}         
\usepackage{times}            

\usepackage{cite}             
\usepackage{tabu}             
\usepackage{booktabs}

\usepackage{hyperref}
\usepackage{url}

\usepackage{siunitx}
\usepackage{amsmath,amssymb,amsfonts}
\usepackage{mathtools, cuted}
\usepackage{algorithmic}
\usepackage[normalem]{ulem}
\usepackage{graphicx}
\usepackage{caption}
\usepackage{subcaption}
\usepackage{xcolor}
\usepackage{float}
\usepackage{tabularx}
\usepackage{array}
\usepackage{balance}
\usepackage{multirow}
\usepackage{svg}

\usepackage{xspace}

\usepackage{fancyhdr}

\newcolumntype{P}[1]{>{\centering\arraybackslash}p{#1}}
\newcolumntype{M}[1]{>{\centering\arraybackslash}m{#1}}

\newenvironment{myitemize}{\begin{list}{$\bullet$}{\renewcommand{\leftmargin}{0.15in}}}{\end{list}}

\onlineid{0}

\vgtccategory{Research}

\title{SiTAR: Situated Trajectory Analysis for In-the-Wild Pose Error Estimation}

\author{Tim Scargill\thanks{e-mail: ts352@duke.edu} %
\and Ying Chen\thanks{e-mail: ying.chen151@duke.edu} %
\and Tianyi Hu\thanks{e-mail: tianyi.hu@duke.edu}
\and Maria Gorlatova\thanks{e-mail: maria.gorlatova@duke.edu}}
\affiliation{\scriptsize Department of Electrical and  Computer Engineering, Duke University}

\teaser{
  \captionsetup[subfigure]{justification=centering}
\begin{subfigure}{.33\textwidth}
  \includegraphics[width=1.0\linewidth]{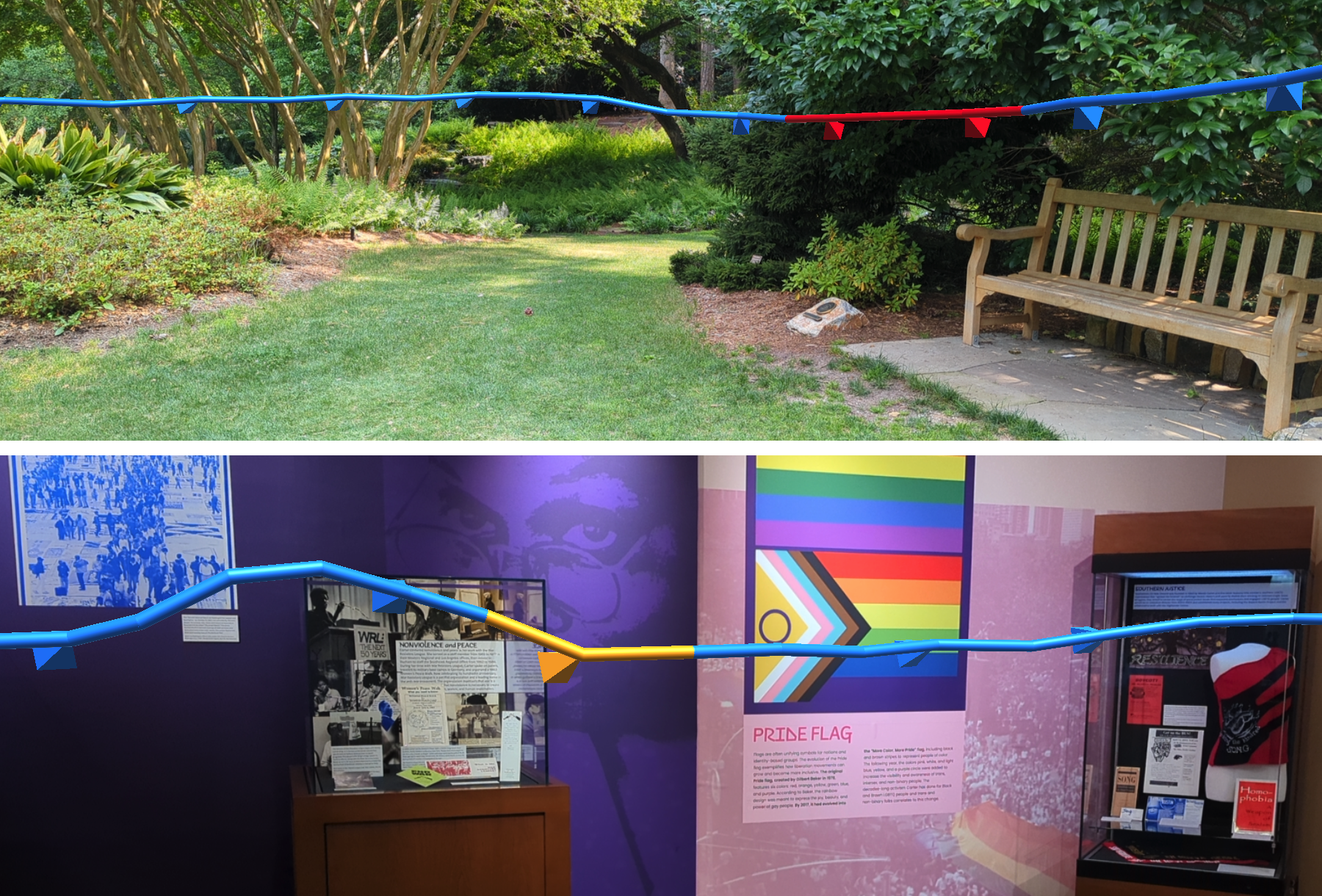}
  \vspace{-0.5cm}
  \caption{}
  \label{fig:VisualizationA}
\end{subfigure}
\hspace{0.05cm}
\begin{subfigure}{.33\textwidth}
  \includegraphics[width=1.0\linewidth]{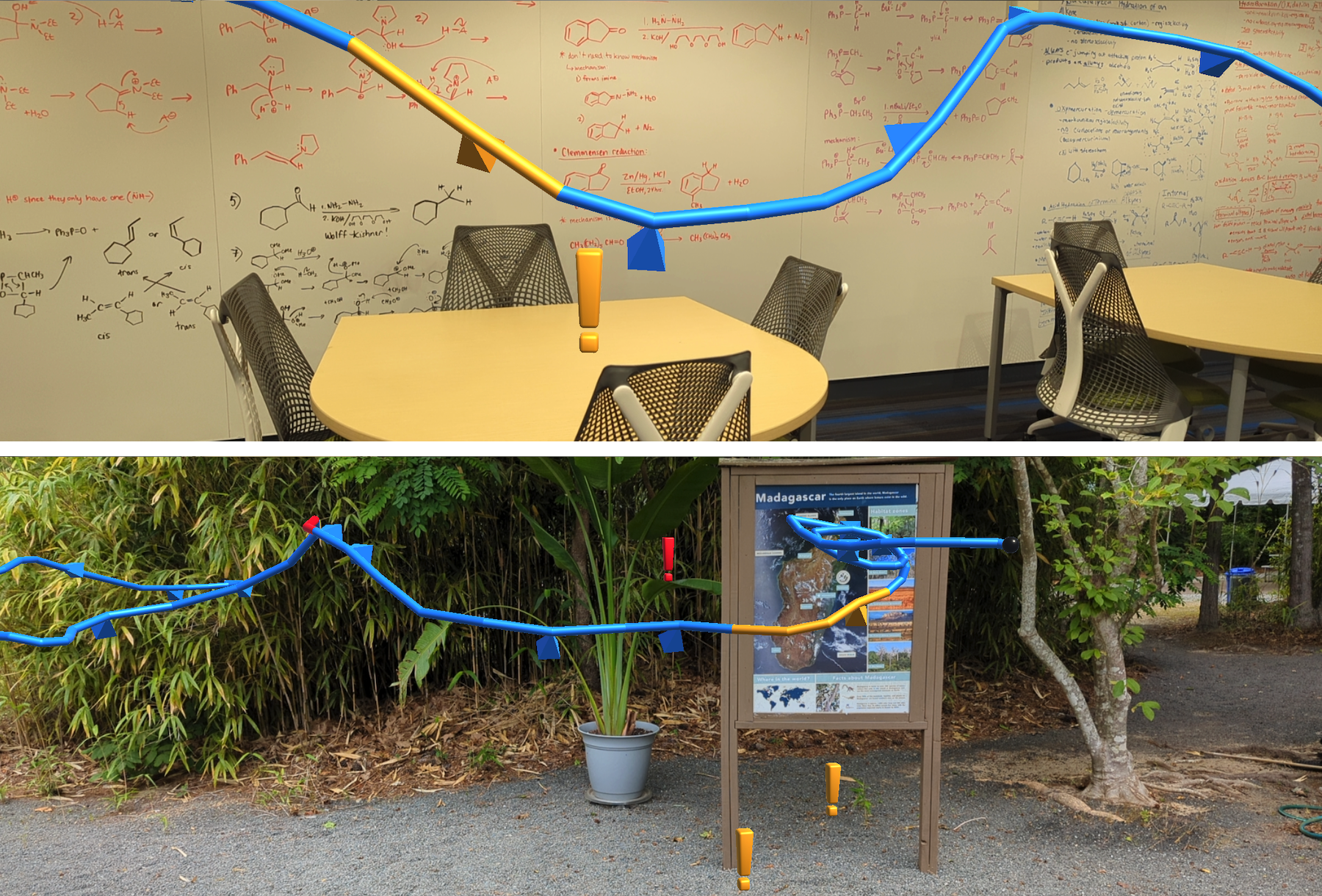}
  \vspace{-0.5cm}
  \caption{}
  \label{fig:VisualizationB}
\end{subfigure}
\hspace{0.05cm}
\begin{subfigure}{.33\textwidth}
  \includegraphics[width=1.0\linewidth]{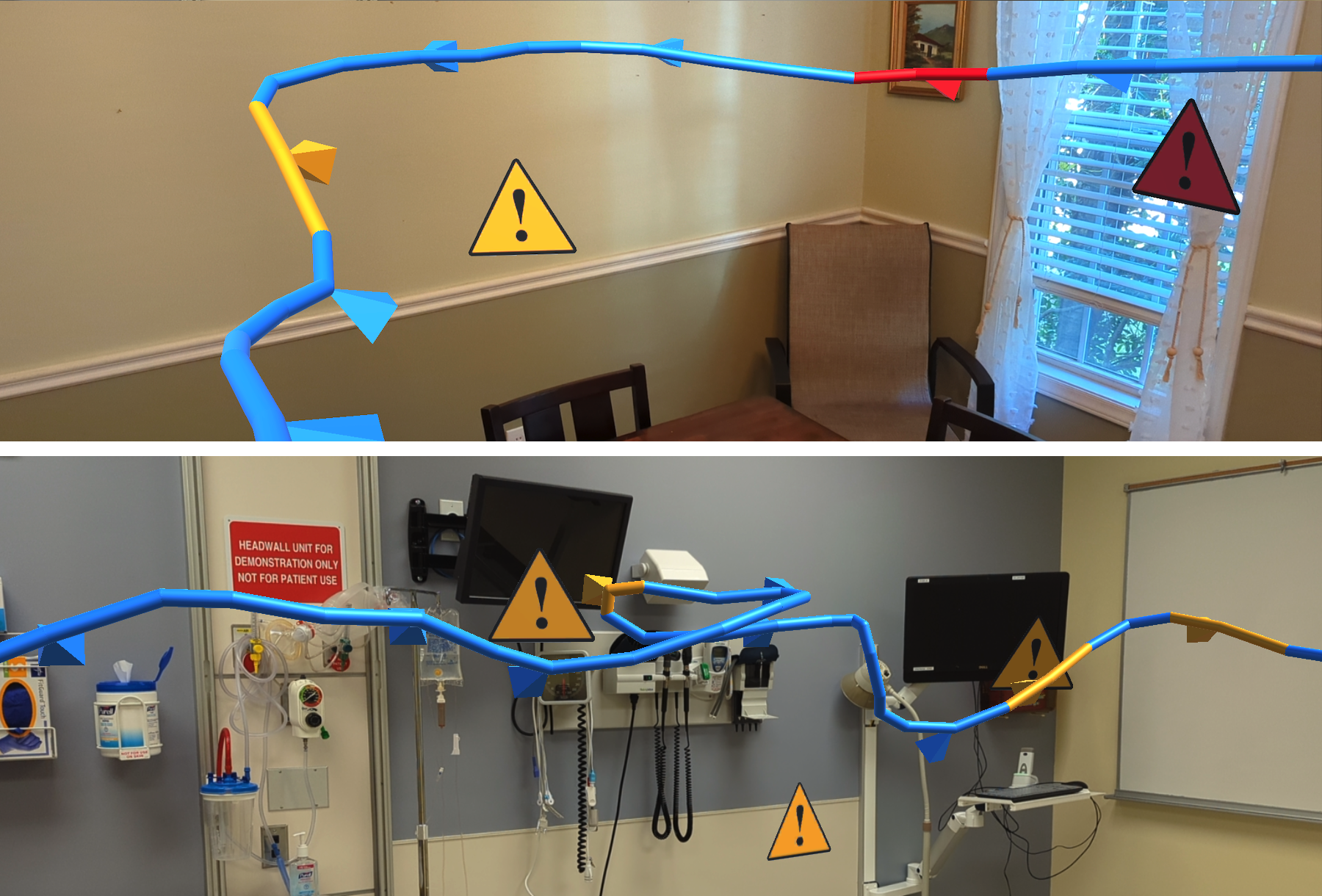}
  \vspace{-0.5cm}
  \caption{}
  \label{fig:VisualizationC}
\end{subfigure}
\vspace{-0.7cm}
\caption{Our situated trajectory analysis system SiTAR in action, identifying environment regions associated with high estimated device pose error. Pictured are examples of environments from our in-the-wild study, in which we tested three visualization techniques: (a) trajectory-only, (b) trajectory + exclamation points, and (c) trajectory + warning signs. Best viewed in color.}
\label{fig:Teaser}
}

\abstract{Virtual content instability caused by device pose tracking error remains a prevalent issue in markerless augmented reality (AR),
especially on smartphones and tablets. However, when examining environments which will host AR experiences, it is challenging to determine where those instability artifacts will occur; we rarely have access to ground truth pose to measure pose error, and even if pose error is available, traditional visualizations do not connect that data with the real environment, limiting their usefulness. To address these issues we present \textbf{SiTAR} (\textbf{\underline{Si}}tuated \textbf{\underline{T}}rajectory Analysis for \textbf{\underline{A}}ugmented \textbf{\underline{R}}eality), the first situated trajectory analysis system for AR that incorporates estimates of pose tracking error. We start by developing the first uncertainty-based pose error estimation method for visual-inertial simultaneous localization and mapping (VI-SLAM), which allows us to obtain pose error estimates without ground truth; we achieve an average accuracy of up to 96.1\% and an average F1 score of up to 0.77 in our evaluations on four VI-SLAM datasets. Next, we present our SiTAR system,
implemented for ARCore devices, combining a backend that supplies uncertainty-based pose error estimates
with a frontend that generates situated trajectory visualizations. Finally, we evaluate the efficacy of SiTAR in realistic conditions by testing three visualization techniques in an in-the-wild study with 15 users and 13 diverse environments; this study reveals the impact both environment scale and the properties of surfaces present can have on user experience and task performance.}

\CCScatlist{
  \CCScatTwelve{Human-centered computing}{Human computer interaction (HCI)}{Interaction paradigms}{Mixed / augmented reality};
}

\fancypagestyle{FirstPage}{
\lfoot{\copyright2023 IEEE. Personal use of this material is permitted. Permission from IEEE must be obtained for all other uses, in any current or future media, including reprinting/republishing this material for advertising or promotional purposes, creating new collective works, for resale or redistribution to servers or lists, or reuse of any copyrighted component of this work in other works.} 
}

\begin{document}

\maketitle

\thispagestyle{FirstPage}

\section{Introduction}
As techniques for persistent augmented reality (AR) mature and become more accessible, we are on the cusp of a long-anticipated future in which virtual content is integrated with the real world across diverse environments. Immersive museum exhibits, navigation guidance in warehouses, factories with virtual displays of sensor data and virtual patient notes in hospitals are just some of the possibilities that can now be realized. However, within this promise still lurks the problem of virtual content instability, caused by AR device pose tracking error. Particularly on handheld devices, the limited field of view of tracking cameras means that challenging environment regions such as blank walls can dominate the visual input data available for the underlying visual-inertial simultaneous localization and mapping (VI-SLAM) algorithm; this results in device pose estimate errors, and virtual content that 
jitters or drifts out of position \cite{ran2019sharear,scargill2022here}.

How then do AR developers decide which parts of an environment are conducive to accurate pose tracking? In other words, \textit{“If I place virtual content here, is it likely to appear stable to users?”} While AR platforms offer general environment design guidelines (e.g., \cite{ARCoreEnvironment}), the complex interaction between lighting, texture, and the geometric and reflective properties of an environment makes estimating the magnitude of errors extremely challenging. Ideally one would quantify pose error by evaluating trajectories against a ground truth pose measurement, but this is infeasible in the vast majority of scenarios due to the cost and logistics associated with obtaining ground truth, e.g., through optical tracking systems \cite{Vicon,OptiTrack}. Finally, some AR platforms provide real-time tracking status \cite{ARCoreTrackingState, ARKitTrackingState}, but they only indicate when tracking is severely compromised or lost completely, and the magnitude of errors is not detected. It is clear we require a pose error estimation method that provides more granular information, and which can be easily deployed in any environment.

Even if accurate pose error estimates are available, another issue remains -- how do we best communicate that information? The pose errors calculated through trajectory evaluations are traditionally reported numerically or in 2D plots (e.g., \cite{campos2021orb, schubert2018tum}), which critically lack any connection with the real environment where a trajectory was created. One might ask, \textit{“There is high error in this part of a trajectory, but which part of the environment does that correspond to? Where was the AR device camera facing at that point?”} This fundamentally limits our ability to understand which parts of an environment are the source of pose estimate errors, and the usefulness of trajectory evaluations for making AR environment design decisions. To fully empower developers to make positive changes that produce high-quality AR experiences, we require not only a method of obtaining accurate pose error estimates, but also a solution that connects those estimates with the environment where a trajectory was created.

To address both of these requirements, we present \textbf{SiTAR} (\textbf{\underline{Si}}tuated \textbf{\underline{T}}rajectory Analysis for \textbf{\underline{A}}ugmented \textbf{\underline{R}}eality), \emph{the first situated trajectory analysis system for AR that incorporates estimates of pose tracking error}. We obtain pose error estimates 
through our development of an uncertainty-based estimation method. Then, to connect those estimates with 
real environment regions 
we employ \mbox{\emph{situated analytics} \cite{elsayed2016situated,thomas2018situated, shin2023reality, fleck2022ragrug,hubenschmid2022relive,luo2023pearl}} in the creation of trajectory visualizations. We test three trajectory visualization techniques for SiTAR in an in-the-wild study, to evaluate it in realistic environments, and the efficacy of our visualizations under diverse conditions. Our key contributions are summarized as follows:

\begin{myitemize}
  \item We develop the first uncertainty-based pose error estimation method for VI-SLAM, which achieves an average accuracy of up to 96.1\% across four datasets, and provides estimates of pose error magnitude \emph{without the need for ground truth pose data}.
  \item We present SiTAR, the first situated trajectory analysis system for AR that incorporates pose error estimates, and implement it for ARCore; 
  we generate uncertainty-based estimates using an edge or cloud backend, and visualize them in the 
  environments they are associated with, on real AR devices. 
  \item We design three visualization techniques for highlighting problematic environment regions associated with high pose error, and test them in an in-the-wild study with 15 participants in 13 diverse environments. Participants indicated a preference for visualizations which attached virtual objects to problematic regions, and our results provide insights into how certain environment properties can impact both task performance and user experience.
\end{myitemize}

The rest of the paper is organized as follows. In Section~\ref{sec:RelatedWork} we cover related work, then in Section~\ref{sec:Uncertainty-basedPoseErrorEstimation} present our uncertainty-based pose error estimation method. We describe our system for situated trajectory analysis on AR devices in Section~\ref{sec:SituatedTrajectoryAnalysisSystem}, develop and evaluate different types of visualization for situated trajectory analysis in  Section~\ref{sec:VisualizationsforSituatedTrajectoryAnalysis}, then summarize our conclusions and future work in Section~\ref{sec:ConclusionsandFutureWork}. The code required for implementation of SiTAR is publicly available at \url{https://github.com/timscargill/SiTAR/}.

\section{Related Work}
\label{sec:RelatedWork}
\textbf{VI-SLAM pose error estimation: }VI-SLAM evaluation tools (e.g., \cite{grupp2017evo, zhang2018tutorial}) are traditionally used to calculate pose error by comparing estimated and ground truth trajectories; this reliance on ground truth pose data is suited to the evaluations conducted with VI-SLAM benchmarks (e.g., \cite{burri2016euroc, schubert2018tum, cortes2018advio, jinyu2019survey, kasper2019benchmark, zuniga2020vi}), but is not scalable to the 
diverse environments which will host AR. On the other hand, AR platforms such as ARCore and ARKit detect when pose estimates are unavailable or questionable without ground truth \cite{ARCoreTrackingState, ARKitTrackingState}, but both only capture severe tracking issues, and do not provide estimates of error magnitude. Recent work has shown promise in the prediction of pose error through the characterization of raw sensor inputs and machine learning \cite{ali2023prediction}, but this only outputs the global consistency of a trajectory; to highlight specific problematic environment regions we require the local consistency, which our 
estimation method provides.

\noindent \textbf{Uncertainty quantification: } 
Existing works on uncertainty quantification across diverse 
fields primarily model data uncertainty~\cite{chang2020data,li2022uncertainty,abdar2021review}, the `noise' and `randomness' inherent in data, and knowledge uncertainty~\cite{UncertaintyEnergy,abdar2021review}, uncertainty due to a system's limited knowledge of input data. The Bayesian-based method through 
posteriori probability formulation 
is a typical uncertainty quantification method~\cite{ping2022statistics}.  
To reduce the computational complexity, Monte Carlo simulation-based \emph{uncertainty propagation} 
has been applied to 
approximate Bayesian results~\cite{zhang2020basic} and quantifies uncertainty by repeated sampling. 
Drawing inspiration yet distinguishing our work from Monte Carlo propagation of data uncertainty induced by measurement errors~\cite{heibetaelmann2019determination}, we are the first to leverage this uncertainty propagation 
methodology to quantify pose estimate uncertainty in SLAM. Here we characterize both data uncertainty (
sensor noise) and knowledge uncertainty (the non-deterministic nature of \mbox{SLAM} algorithms). Prior works have quantified the uncertainty of the optimization-based SLAM backend for entire trajectories using the maximum likelihood pose estimator covariance matrix, to guide decision-making in SLAM (e.g., \cite{khosoussi2019reliable, carlone2018attention, chen2021anchor, chen2023adaptslam}); our work differs in that we capture local uncertainty by considering individual sub-trajectories, and that we use the pose estimate \emph{output} to quantify uncertainty, such that our method can be applied to both open-source SLAM 
and the `black-box' algorithms on commercial AR platforms. 

\noindent \textbf{Situated analytics and semantic trajectories: }We propose and develop a new application for situated analytics \cite{elsayed2016situated,thomas2018situated, shin2023reality, fleck2022ragrug,hubenschmid2022relive,luo2023pearl}, 
data analysis 
supported by \emph{situated visualizations} -- for a recent review see \cite{bressa2021s}. 
Specifically, our system semantically enriches AR device trajectory visualizations with pose error estimates; this inclusion of data beyond pose in trajectory visualization is related to 
work on \emph{semantic trajectories} \cite{parent2013semantic, yan2013semantic, albanna2015semantic}. One example is 
coloring 
2D trajectory plots according to pose error magnitude in SLAM trajectory evaluations \cite{grupp2017evo}, which we build on to produce 3D, situated trajectory visualizations in AR. Previous works have demonstrated visualizing various types of trajectories in AR, including the planning or communicating of robot or drone movements \cite{zollmann2014flyar,quintero2018robot,walker2018communicating,chen2021pinpointfly} and guidance for navigation \cite{gay2010augmented, pankratz2013user}, medical procedures \cite{zhao2020intelligent, eom2022neurolens, condino2020wearable}, and sports training \cite{lin2021towards}. Most closely related is \cite{buschel2021miria}, a recent toolkit that provides AR-based situated analytics for AR user movement and interactions. However, to the best of our knowledge our system is the first to enable situated analytics for device pose error.

\begin{figure}
\centering
\captionsetup[subfigure]{justification=centering}
\begin{subfigure}{.23\textwidth}
  \includegraphics[width=1\linewidth]{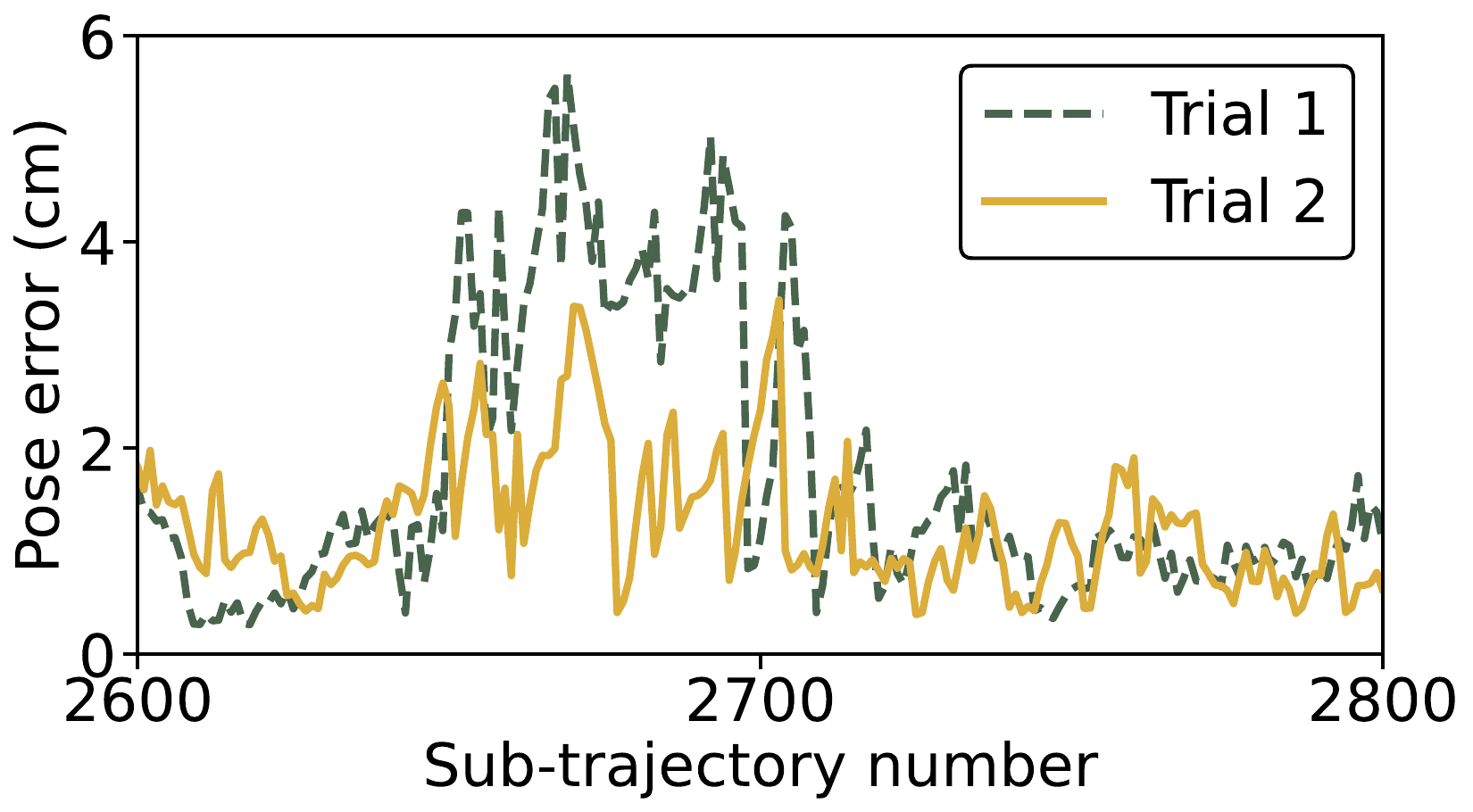}
  \vspace{-0.6cm}
  \caption{Pose error diversity in two trials}
  \label{fig:STA1_trials}
\end{subfigure}
\begin{subfigure}{.23\textwidth}
  \includegraphics[width=1\linewidth]{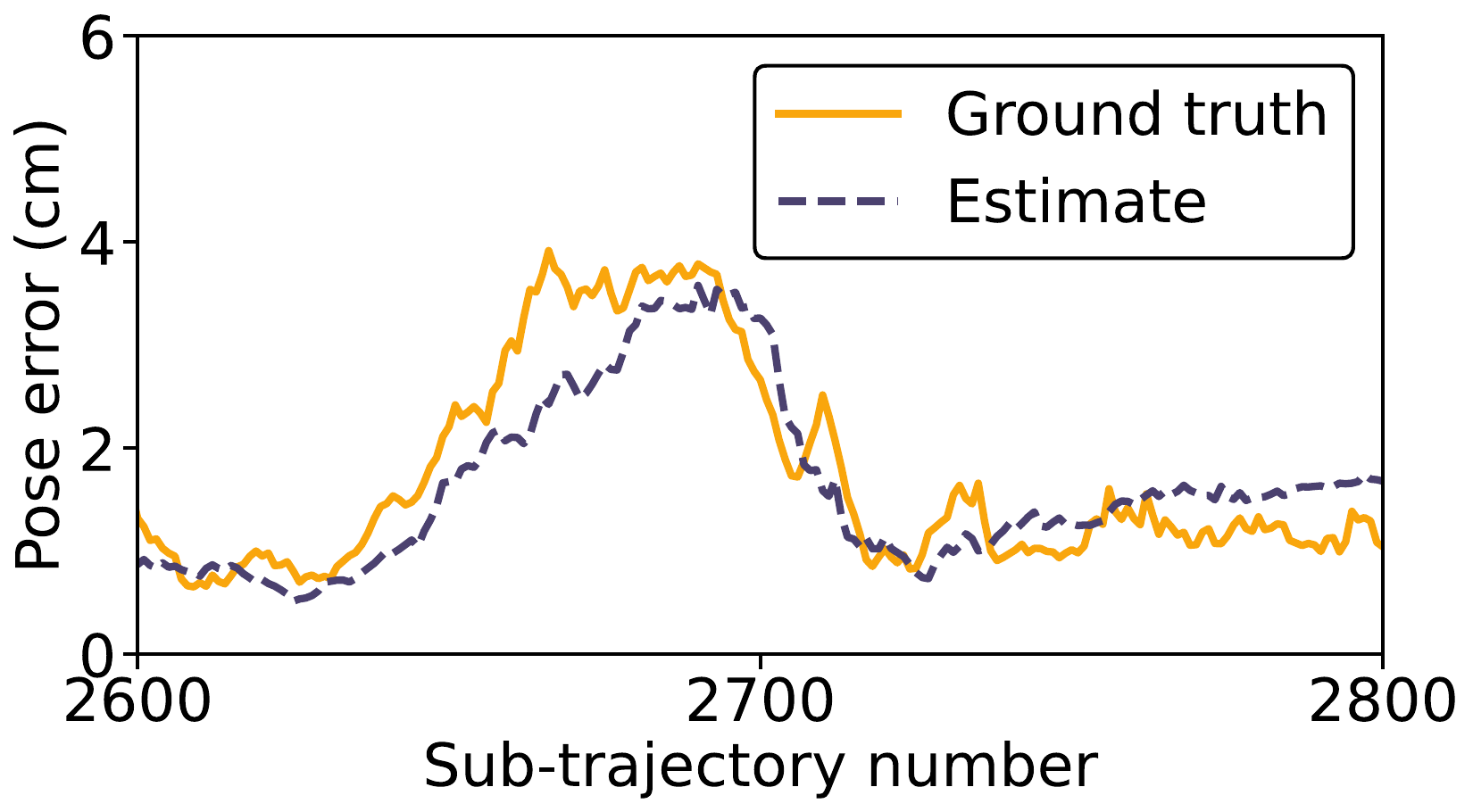}
  \vspace{-0.6cm}
  \caption{Example of our pose error estimates}
  \label{fig:STA1_estimation}
\end{subfigure}
\vspace{-0.3cm}
\caption{Our uncertainty propagation-based pose error estimation uses (a) the diversity of pose estimates in repeated trials to (b) estimate the average ground truth-based pose error for that input data. 
Samples shown are from the SenseTime A1 sequence \cite{jinyu2019survey}.}
\label{fig:STA1_Examples}
\vspace{-0.6cm}
\end{figure}

\section{Uncertainty-based Pose Error Estimation}
\label{sec:Uncertainty-basedPoseErrorEstimation}
\subsection{Background and Motivation}
Virtual content stability in markerless AR is determined by the accuracy of pose estimates achieved through a VI-SLAM algorithm. However, errors are often present in VI-SLAM-based pose estimates due to challenging visual or inertial input data \cite{jinyu2019survey}, and as a result virtual content instability is a known issue in AR \cite{ran2019sharear,scargill2022here}. The magnitude of 
errors is challenging to predict; not only can multiple different input data properties cause errors, but the inner VI-SLAM pipeline has multiple elements, and errors can occur at various points. Furthermore, relationships between input data characteristics and pose error obtained in one environment do not generalize to new environments, due to the complex interactions between
lighting, textures, and the geometric and reflective properties of objects present. 

The core motivation for our uncertainty-based 
estimation approach is the stochastic nature of pose estimates in state-of-the-art VI-SLAM (e.g., \cite{campos2021orb, qin2018vins}). We illustrate this 
in Figure~\ref{fig:STA1_trials} by plotting the translational pose error in each 10-frame sub-trajectory used to calculate relative error \cite{zhang2018tutorial} 
when running the A1 sequence from the SenseTime 
benchmark \cite{jinyu2019survey} in ORB-SLAM3 \cite{campos2021orb}. 
Different trials result in varying 
error magnitudes, and this variance -- pose estimate uncertainty -- is generally greater when 
error is higher. Intuitively, this reflects our observation that \emph{accurate trajectories are alike, whereas inaccurate trajectories are wrong in their own way}. This 
randomness arises due to the use of RANSAC \cite{fischler1981random} in SLAM to select a set of good feature matches, inliers, between neighboring camera images; because RANSAC uses repeated random sub-sampling, and the number of iterations is limited due to 
real-time operation, it can produce sub-optimal results, particularly when the proportion of outliers is high (i.e., input data are challenging) \cite{raguram2008comparative, noll2010real}.

\subsection{Pose Error Estimation Method}
\label{subsec:Uncertainty-basedEstimationMethod}
Our pose error estimation method is based on the propagation of uncertainty, the effect that the randomness or uncertainty of input variables has on the system output. Uncertainty propagation has been used to quantify output uncertainty in various domains, from transistor models \cite{petrocchi2017measurement} to visual analytics \cite{sacha2015role}. 
Specifically, we quantify how the randomness inherent in RANSAC-based inlier selection affects the uncertainty of VI-SLAM pose estimates by performing repeated trials with the same visual and inertial input data, similar to works that use Monte Carlo simulation for uncertainty propagation (e.g., \cite{madsen2006methods}). To support repeated trials in our system for AR devices we employ an edge or cloud server and additional AR devices (see Section~\ref{sec:SituatedTrajectoryAnalysisSystem}). Moreover, we show that this uncertainty, i.e., the diversity of pose estimates, can be used to estimate the average pose error that results from a set of visual and inertial input data, and thereby identify environment regions likely to result in high pose error.

Pose estimates generated in repeated SLAM trials are not necessarily in the same coordinate frame, which means that standard precision metrics cannot be applied to these unaligned estimates directly. Therefore, to calculate pose estimate diversity across a set of repeated trials, we perform pairwise trajectory evaluations (which include the alignment step) on the trajectory estimates obtained from those trials, for all possible trial pairs, such that for $n$ trials we perform $n(n-1)/2$ evaluations.  We use $B$ to denote this set of $n$ trajectory estimates. We denote each trajectory estimate as $\mathbf{b}_i= ( {b_1^i,b_2^i, \cdots ,b_S^i} )\in B$, where $i$ ranges from 1 to $n$, and $S$ represents the total number of sub-trajectories within a trajectory estimate.  
 In the trajectory evaluations we calculate \emph{relative pose error}, a measure of local trajectory consistency \cite{kummerle2009measuring, zhang2018tutorial}, because this allows us to associate pose error with sub-trajectories, smaller portions of the visual and inertial input data, and hence localized areas of the environment. More formally, we run trajectory evaluation $E$ on all possible pairs $\{\mathbf{b}_i,\mathbf{b}_j\}$ of trajectory estimates in  $B$, 
 and obtain the set $e_B$ of pose errors, i.e., $e_B = \{ { E(\mathbf{b}_i,\mathbf{b}_j) |\mathbf{b}_i,\mathbf{b}_j \in B, i \ne j } \}$. 
For each sub-trajectory we capture translational pose error, on the basis that errors in virtual object position are more noticeable than orientation errors, but this method could equally be applied to rotational pose error. Thus, for each sub-trajectory we have $n(n-1)/2$ translational pose errors which represent pose output diversity. Rewriting $E(\mathbf{b}_i,\mathbf{b}_j)$ as $E(\mathbf{b}_i,\mathbf{b}_j) \coloneqq (e^{i,j}_1,e^{i,j}_2,\cdots,e^{i,j}_S)$, for the $k$-th sub-trajectory, these $n(n-1)/2$ 
pose errors constitute the set $\{{e_k^{i,j}} |{\mathbf{b}_i},{\mathbf{b}_j} \in B, i\ne j\}$.

To obtain our final pose error estimate for each sub-trajectory, we then apply a statistical measure $M$ to characterize pose output diversity. When evaluating our method (Section~\ref{subsec:Uncertainty-basedEstimationEvaluation}) we test various statistical measures as methods of estimating pose error from diversity, which we term \textit{uncertainty calculation settings}. These include the mean, median, trimmed means (the mean calculated after a predefined percentage of the smallest and largest values are removed from the data), and the mean after outliers are removed.

Formally, applying our uncertainty propagation-based method to trajectory estimates in 
$B$, we derive the pose error estimate $
U_{B,k}
$ for the $k$-th sub-trajectory as
$U_{B,k} = M(\{{e_k^{i,j}} |{\mathbf{b}_i},{\mathbf{b}_j} \in B, i\ne j\})$.
 We then posit that for 
 a set of trajectory estimates $A$, which is disjoint from $B$ and obtained from the same inputs as $B$,
 the average pose error $\overline{e_A}$ can be estimated as  $\overline{e_A} = 
(U_{B,1},U_{B,2},\cdots,U_{B,S})
$.
  An example of the results we obtain using this method is shown in Figure~\ref{fig:STA1_estimation}, which shows the average pose error from 100 trials of the SenseTime A1 sequence using ORB-SLAM3 (labeled `Ground truth') along with our estimate, calculated using 100 separate trials and a 30\% trimmed mean. To convert numerical pose error estimates to a form easily interpretable by AR users, one can classify them by error magnitude, e.g., as low, medium or high error for each sub-trajectory -- we apply this type of classifier in Section~\ref{subsec:Uncertainty-basedEstimationEvaluation} to evaluate the performance of our pose error estimation method.



\subsection{Pose Error Estimation Evaluation}
\label{subsec:Uncertainty-basedEstimationEvaluation}
\subsubsection{VI-SLAM Datasets}
\label{subsubsec:VI-SLAMDatasets}
To the best of our knowledge there is no existing method for estimating relative pose error without ground truth pose measurements. Therefore, we evaluate our estimates against pose error calculated using ground truth pose, on 30 sequences from four VI-SLAM datasets. We select two widely-used datasets with motion patterns representative of handheld AR devices: from TUM VI \cite{schubert2018tum} we use all six `room' sequences \mbox{(room1-6)}, and from SenseTime we use all eight `A' sequences \mbox{(A0-7)}. We then increase the diversity of visual environments 
using our recent method for combining inertial data from existing SLAM datasets with visual data from virtual environments \cite{scargill2022integrated}.
We create two new VI-SLAM datasets using this method: `Hall' and `LivingRoom'. For Hall we generate eight sequences which combine the
SenseTime inertial data with visual data from a sample virtual environment in Unity's High Definition Rendering Pipeline (HDRP) \cite{HDRP}, a large 11m$\times$7m$\times$4m space with multiple different light sources, reflective surfaces and a curved wall. For LivingRoom we generate eight sequences using the SenseTime trajectories and a virtual environment we created in Unity's HDRP, an 8m$\times$6m$\times$4m room containing both highly textured (e.g., a painting, soft furnishings) and featureless regions (a blank wall). Our Hall and LivingRoom datasets are publicly available to download at \url{https://github.com/timscargill/SiTAR/}. 

\subsubsection{Evaluation Setup}
\label{subsubsec:Uncertainty-basedEstimationEvaluationSetup}
For our main evaluation we run all 30 sequences with a current state-of-the-art VI-SLAM algorithm, ORB-SLAM3 \cite{campos2021orb} (monocular version, default settings). For an additional comparison we run the TUM VI sequences with another state-of-the-art VI-SLAM algorithm, VINS-Mono \cite{qin2018vins} (with a RANSAC threshold of 5px, which provides a good balance between accuracy and precision \cite{riu2022classification}). In our main evaluation we perform, for each sequence, 100 trials to calculate average ground truth-based pose error and 100 separate trials which we use to generate our pose error estimates. In our evaluation with VINS-Mono we perform 10 trials for ground truth-based error and 10 separate trials for estimation. We use a desktop computer (Intel i7-9700K CPU, Nvidia GeForce RTX 2060 GPU), but run SLAM on a virtual machine with 4 CPUs and 8GB RAM, representative of the computational resources of a mobile AR device. We perform trajectory evaluations using the evo Python package \cite{grupp2017evo}, with 10-frame sub-trajectories. Our testing indicated that 10 frames is a good balance between limiting visual input data to one environment region while maintaining sufficient error variance for informative results; the selection of optimal sub-trajectory lengths for different environments is an interesting topic for future work.

We test our 
estimation method with seven uncertainty calculation settings: the \textit{mean}, \textit{median}, trimmed means at 10\%, 20\%, 30\%, 40\% (\textit{trim\_mean\_10}, \textit{trim\_mean\_20}, \textit{trim\_mean\_30}, \textit{trim\_mean\_40}), and the mean after removing outliers defined by the interquartile range (\textit{mean\_inliers}). For each setting we record results when the first 5, 10, 20, 30, 40, 50, 60, 70, 80, 90 and 100 trials are used to calculate estimates. We evaluate our estimates with the performance of three 3-class error classifiers, each with different class boundaries: 5cm \& 10cm; 2cm \& 10cm; 2cm \& 5cm. We choose these magnitudes because when applied to the ground truth data, they produce the imbalanced distribution we desire in our outputs; \emph{to provide actionable information, our system should present a small number of sub-trajectories 
with especially high pose error, rather than an even distribution across all classes}. We quantify estimation
performance using average accuracy and F1 score across all sequences.

\begin{figure}
\centering
\captionsetup[subfigure]{justification=centering}
\begin{subfigure}{.45\textwidth}
  \includegraphics[width=1\linewidth]{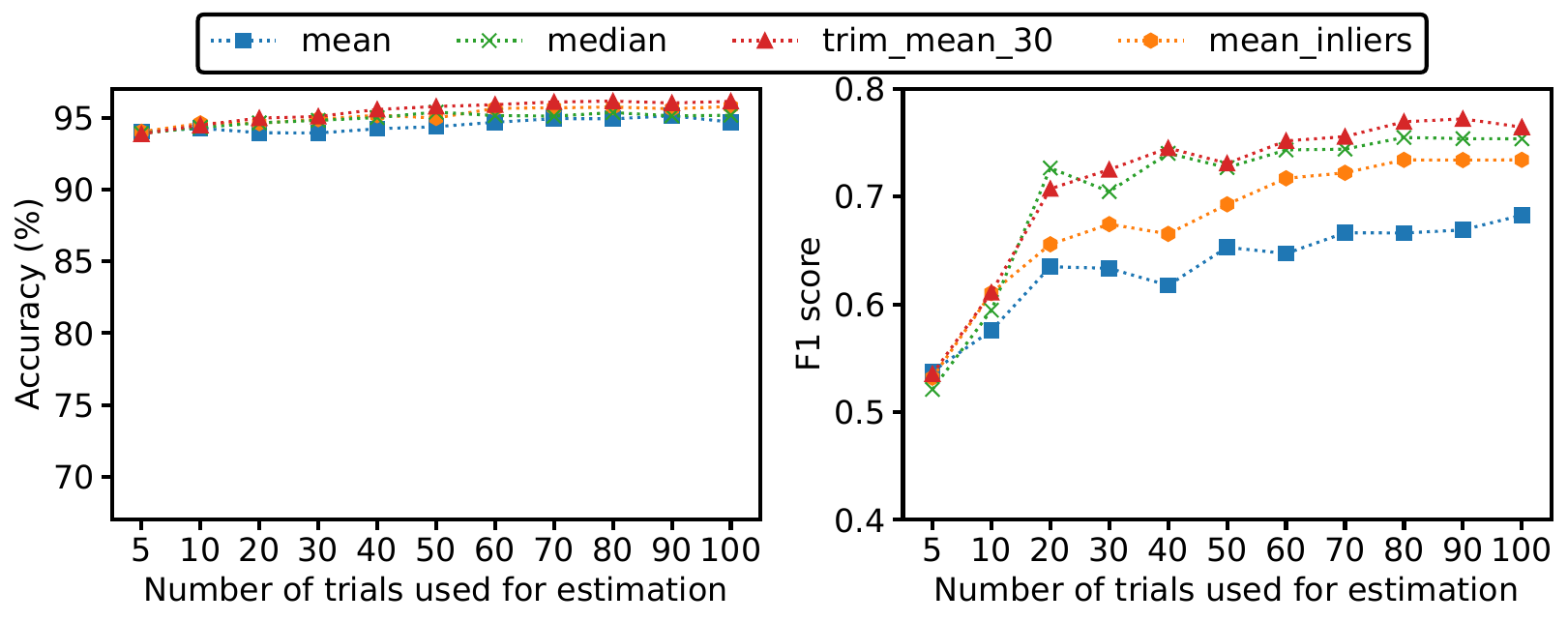}
  \vspace{-0.6cm}
  \caption{Accuracy (left) and F1 scores (right) with class boundaries of 5cm \& 10cm}
  \label{fig:Classifier1}
\end{subfigure}
\begin{subfigure}{.45\textwidth}
  \includegraphics[width=1\linewidth]{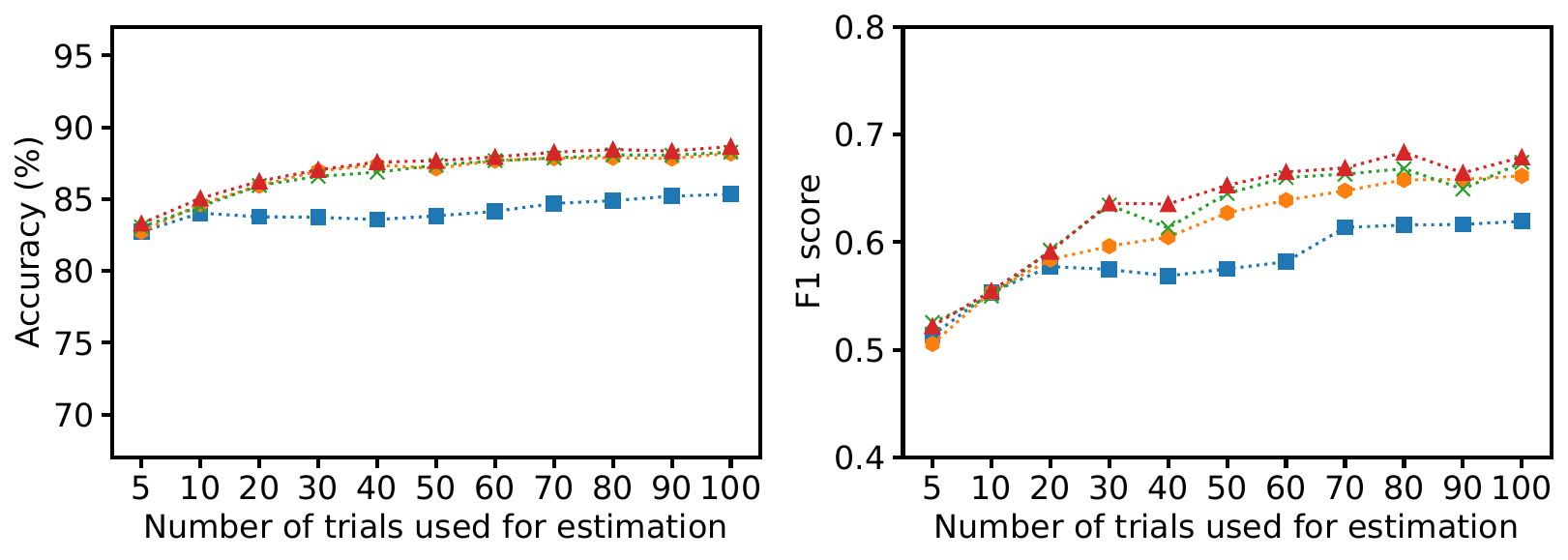}
  \vspace{-0.6cm}
  \caption{Accuracy (left) and F1 scores (right) with class boundaries of 2cm \& 10cm}
  \label{fig:Classifier2}
\end{subfigure}
\begin{subfigure}{.45\textwidth}
  \includegraphics[width=1\linewidth]{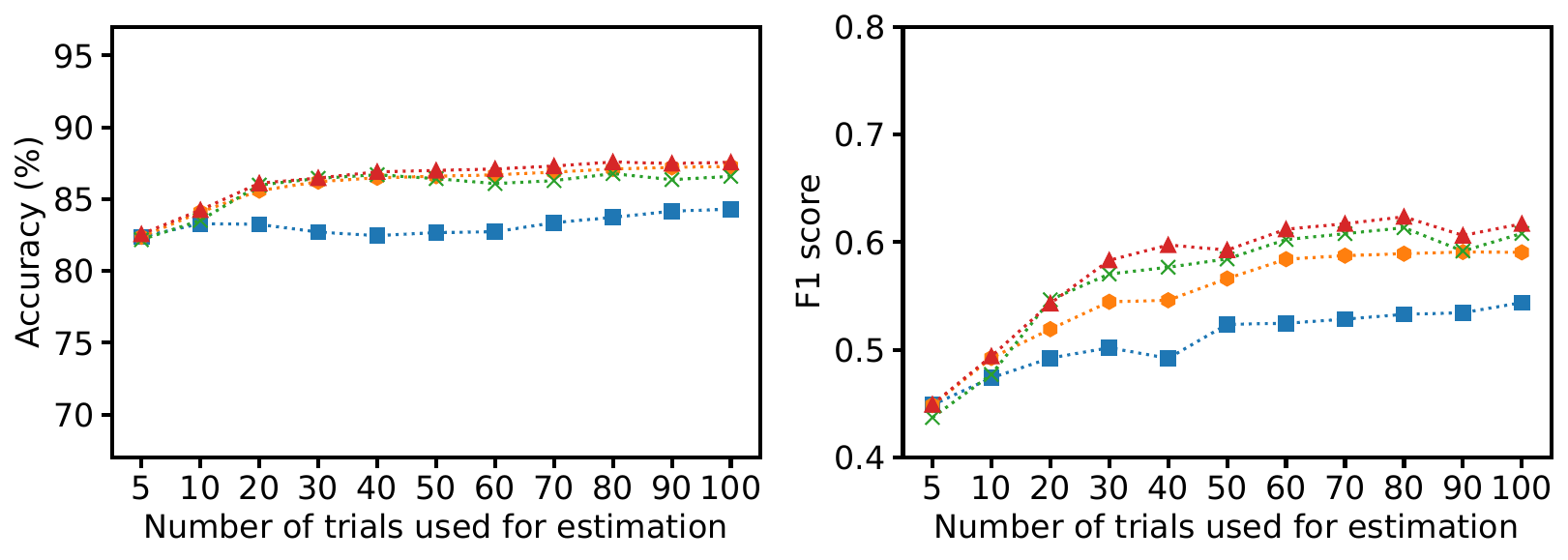}
  \vspace{-0.6cm}
  \caption{Accuracy (left) and F1 scores (right) with class boundaries of 2cm \& 5cm}
  \label{fig:Classifier3}
\end{subfigure}
\vspace{-0.3cm}
\caption{Performance of our uncertainty-based pose error estimation method, using different settings (see legend) and different numbers of trials. Applying outlier removal when calculating estimates (e.g., \textit{trim\_mean\_30}) outperformed the standard \textit{mean}.
}
\label{fig:UncertaintyEstimationResults}
\vspace{-0.7cm}
\end{figure}

\subsubsection{Evaluation Results}
\label{subsubsec:Uncertainty-basedEstimationEvaluationResults}
Results for our main evaluation are shown in Figure~\ref{fig:UncertaintyEstimationResults}; for each class boundary setting (Figure~\ref{fig:Classifier1}: 5cm \& 10cm, Figure~\ref{fig:Classifier2}: 2cm \& 10cm, Figure~\ref{fig:Classifier2}: 2cm \& 5cm) we plot accuracy (left) and F1 score (right) with different numbers of trials used for estimation. Each line plot represents an uncertainty calculation setting (see legend) from Section~\ref{subsec:Uncertainty-basedEstimationMethod}; for readability we only include the best-performing trimmed mean, omitting \textit{trim\_mean\_10} (similar results to \textit{mean\_inliers}), \textit{trim\_mean\_20} (similar to \textit{trim\_mean\_30}), and \textit{trim\_mean\_40} (similar to \textit{median}). Our estimation method achieves a peak accuracy of 96.1\% (\textit{trim\_mean\_30}, 80 trials), and a peak F1 score of 0.77 (\textit{trim\_mean\_30}, 90 trials), both for the class boundary setting of 5cm \& 10cm. 
Accuracy and F1 scores were lower for the other more challenging class boundary settings, but performance is still sufficient to provide a useful output. Higher accuracy and F1 scores were generally achieved as more trials were used for estimation, though there was a drop in F1 scores at 40 trials with the \textit{mean}, due to large outliers between trials 31 and 40 causing over-prediction of error for three sequences. 30 trials resulted in comparable performance to 100 for all class boundary settings (with \textit{trim\_mean\_30} the maximum accuracy drop was 1.6\%, and the maximum drop in F1 score was 0.04), indicating this may provide a good balance between estimation performance and latency overhead 
(see Section~\ref{subsec:SystemLatencyCharacterization}).

When we examine the performance of different uncertainty calculation settings, 
\textit{trim\_mean\_30} achieves the highest accuracy and F1 score for all class boundary settings. The \textit{median} (essentially trimming at 50\%) yields competitive performance, but we lose some valuable information. \textit{Mean\_inliers} removes only a small percentage of outliers, and F1 scores suffer as a result. Notably the \textit{mean}, which includes all error values, performs the worst of all settings with more than five trials. For example, when estimating using 40 trials, with class boundaries of 5cm \& 10cm, \textit{trim\_mean\_30} achieves an accuracy of 95.6\% and an F1 score of 0.75, while the \textit{mean} achieves 94.2\% and 0.62 respectively. This highlights the benefit of outlier removal, especially when using larger numbers of trials. 

Interestingly, this performance boost from outlier removal is not apparent in the sensitivity and specificity values for each uncertainty calculation and class boundary setting, shown in Table~\ref{tab:SensitivitySpecificity} (calculated individually for each class then averaged). In fact, sensitivity and specificity are greater for the \textit{mean} because they do not take into account the prevalence of different classes \cite{erickson2021magician}. For sequences with a small number of high error instances the \textit{mean} results in several false positives for the high error class, but this does not lower specificity greatly, due to the large number of true negatives. However, if outlier-removing methods such as \textit{trim\_mean\_30} fail to detect the small number of true positives, this results in low sensitivity for the high error class. We are more likely to detect all high error 
regions with the \textit{mean}, but it is also more likely to result in warnings for `good' regions, which would reduce the usefulness of our solution.     

In contrast, in our evaluation with the VINS-Mono the best-performing uncertainty calculation setting was the \textit{mean}; here our estimation method achieved an average accuracy of 99.4\% and an average F1 score of 0.70 for class boundaries of 5cm and 10cm, 95.2\% and 0.57 respectively for class boundaries of 2cm and 10cm, and 95.0\% and 0.52 respectively for class boundaries of 2cm and 5cm. This difference in optimal uncertainty calculation setting for VINS-Mono is due to different pose error distributions across multiple trials compared to ORB-SLAM3, and suggests that our pose error estimation method may benefit from being configured for the specific VI-SLAM algorithm employed on an AR platform.

\begin{table}
  \centering
  \caption{Mean specificity and sensitivity in our pose error estimation evaluation (100 trials), for each
  class boundary setting (5cm \& 10cm, 2cm \& 10cm, 2cm \& 5cm) and uncertainty calculation setting.} 
  \vspace{-0.3cm}
  \begin{tabular}{|P{1.7cm}||P{0.6cm}|P{0.6cm}|P{0.6cm}|P{0.6cm}|P{0.6cm}|P{0.6cm}|}
  \hline
 & \multicolumn{3}{c|}{\textbf{Sensitivity}} & %
    \multicolumn{3}{c|}{\textbf{Specificity}} \\
\cline{1-7}
 \textbf{Setting} & \textbf{5-10} & \textbf{2-10} & \textbf{2-5} & \textbf{5-10} & \textbf{2-10} & \textbf{2-5} \\
\hline
\textit{mean}&0.87&0.75&0.72&0.88&0.80&0.82\\
\hline
\textit{median}&0.86&0.70&0.66&0.81&0.72&0.75\\
\hline
\textit{trim\_mean\_30}&0.87&0.70&0.67&0.83&0.73&0.76\\
\hline
\textit{mean\_inliers}&0.87&0.73&0.69&0.84&0.76&0.79\\
\hline
\end{tabular}
  \label{tab:SensitivitySpecificity}
  \vspace{-0.6cm}
\end{table}

\subsubsection{Evaluation Discussion}
\label{subsubsec:Uncertainty-basedEstimationEvaluationDiscussion}
In a statistical context, 
research has demonstrated the disproportionate impact of outliers on analyses that rely on the sample mean and variance; for example, outliers can undermine 
the efficacy of parametric tests such as $t$-tests and $F$-tests~\cite{liao2016outlier}. 
The benefit of outlier removal increases when estimating pose error through uncertainty propagation because of the method used to calculate pose estimate uncertainty, pairwise trajectory evaluations. 
This is clear when we examine the relative contribution of each trial by calculating the ratio of error values that a single trial impacts to the total number of error values. For the ground truth-based data this is $1/n$, where $n$ is the total number of trials. However, for pairwise comparisons this is the number of other trials which one trial can be compared with, $n-1$, over the total number of trials $n(n-1)/2$, giving us $2(n-1)/[n(n-1)] = 2/n$. This increased influence of outliers in pairwise comparisons explains the need to apply statistical measures such as the trimmed mean to remove extreme error values.  

Our pose error estimation method is predicated on the observation that inaccurate trajectories are wrong in their own way. We analyzed our estimation performance for all 30 sequences individually and found that while this holds true for the majority, there are three for which it does not, 
resulting in noticeably worse performance. For example, despite SenseTime A0 being a challenging sequence that results in high pose error \cite{jinyu2019survey}, trajectory estimates are similar; here inaccurate trajectories are wrong in a consistent way, such that our uncertainty-based estimates are lower than the actual pose error. Interestingly, this is also the case for Hall A0 and LivingRoom A0, suggesting the possible influence of the type of trajectory or inertial data. The underlying reason for these inaccurate yet consistent trajectory estimates is a topic we will examine in future work.



\section{Situated Trajectory Analysis System: SiTAR}
\label{sec:SituatedTrajectoryAnalysisSystem}
In this section we describe the motivation, design and architecture for SiTAR, our situated trajectory analysis system for AR that provides visualizations of uncertainty-based pose error estimates. 


\subsection{Motivation}
\label{subsec:Motivation}
Traditionally, evaluations of device pose tracking accuracy are presented through numeric error values and 2D plots of ground truth and estimated trajectories (e.g., \cite{campos2021orb, schubert2018tum, jinyu2019survey}). While essential for benchmarking tracking algorithms, neither communicates the environment regions visible through the device camera when errors occur. This is vital information for AR installation designers because pose error causes spatially registered virtual content in the current camera view to appear unstable to users. 
We require a solution that allows designers to identify the environment regions visible when pose tracking errors occur, so that they can either avoid placing virtual content there, or adjust environment properties to support better tracking. 


To this end, we propose the use of \emph{situated visualizations} for trajectory analysis, such that trajectory evaluation results are displayed relative to the real environment in which the trajectory was created. For example, a 3D trajectory can display where an AR device moved, while other types of visualization can be rendered at the locations visible in device camera views. These visualizations can then be semantically enriched with indicators of estimated pose error magnitude, such as color. Connecting trajectory evaluation results to the environment in this manner will allow AR designers and developers to quickly identify problematic environment regions, without relying on generic qualitative guidelines or resorting to laborious methods such as matching estimated error values with recorded camera images through timestamps. This empowers them to place virtual content with confidence, to rapidly implement and test environment optimization techniques, and opens up exciting possibilities for integrating predictive analytics into AR experience design.

\begin{figure}
\includegraphics[width=1\columnwidth]{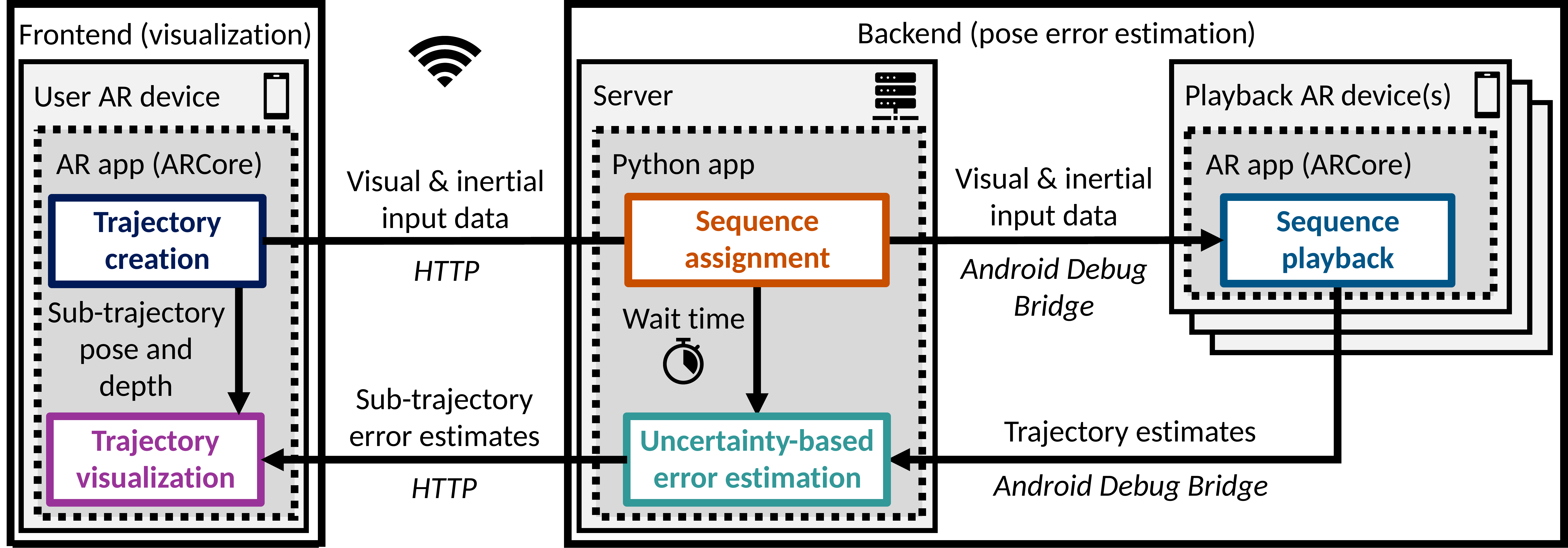}
\vspace{-0.24in}
\caption{
SiTAR system architecture for situated trajectory analysis on AR devices using
uncertainty
-based pose error estimation. 
}
\label{fig:SystemArchitecture}
\vspace{-0.65cm}
\end{figure}

\subsection{System Design}
\label{subsec:SystemDesign}
Our uncertainty propagation-based pose error estimation method (Section~\ref{subsec:Uncertainty-basedEstimationMethod}) requires the replaying of sequences (visual and inertial input data) multiple times to obtain multiple trajectory estimates; because ARCore provides native support for this through their Recording and Playback API \cite{RecordingandPlayback} we describe the implementation of SiTAR for ARCore here, but this can be adapted for any AR platform which facilitates the recording and playback of sequences.\hfill \break

\begin{figure}
\centering
\includegraphics[width=0.7\linewidth]{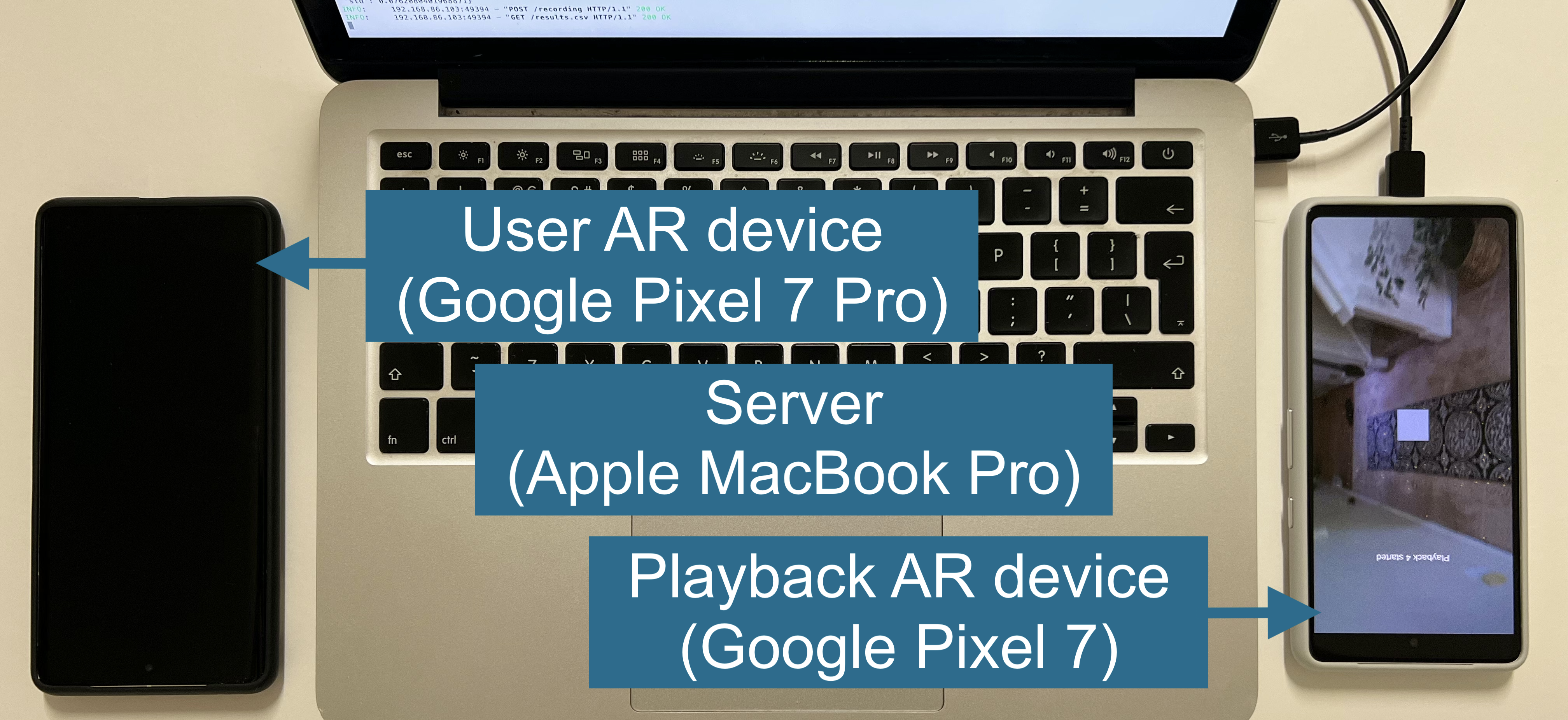}
\vspace{-0.24cm}
\caption{An example SiTAR hardware setup for ARCore, with an Apple MacBook Pro used as an edge server for the system backend.}
\label{fig:SystemImage}
\vspace{-0.55cm}
\end{figure}

\noindent \textbf{System overview: }The SiTAR system architecture is shown in Figure~\ref{fig:SystemArchitecture}. SiTAR consists of a frontend which handles trajectory creation and visualization on the user AR device, and a backend which provides pose error estimates to enrich trajectory visualizations. The system backend comprises a server and one or more playback AR devices; separate playback AR devices are employed because sequences must be replayed in real time, and using them frees up the user's AR device in the meantime. Furthermore, the use of multiple playback AR devices enables the parallelization of sequence playback, so multiple trajectory estimates can be obtained in less time. Our system backend can be deployed using either an edge or cloud server, and with physical playback AR devices or Android emulators. 
Figure~\ref{fig:SystemImage} shows an example hardware setup for SiTAR. Next, we describe the function of each system module in Figure~\ref{fig:SystemArchitecture}, in the order corresponding to the main flow of data, i.e., \mbox{Trajectory creation} $\rightarrow$ Sequence assignment $\rightarrow$ Sequence playback $\rightarrow$ Uncertainty-based error estimation $\rightarrow$ \mbox{Trajectory visualization}.

\noindent \textbf{Trajectory creation: }The trajectory creation module allows users to create a trajectory (visual and inertial input data plus timestamped pose estimates) in an AR app on an AR device. We created this app in Unity 2021.3.14f1, using AR Foundation 4.2. Users start a trajectory with a UI button press, model AR user device movements, then end the trajectory with a button press. During the trajectory we capture visual and inertial input data using the ARCore Recording and Playback API, which saves the data in .mp4 format. The estimated position and orientation of the AR device are logged every five camera frames, such that the endpoints and midpoints of each 10-frame sub-trajectory are captured. To inform the rendering of certain visualizations (see Section~\ref{subsec:VisualizationTechniques}) we also record depth estimates using the ARCore Depth API \cite{DepthAPI}, at the sub-trajectory midpoints. Once a trajectory has been created, the visual and inertial input data are transmitted to the server via an HTTP POST request.

\noindent \textbf{Sequence assignment: }The sequence assignment module is implemented in a Python app on the server, built using the FastAPI \cite{FastAPI} framework. This module handles an HTTP POST request containing the visual and inertial input data -- a \textit{sequence} -- for a created trajectory, and saves these data to local storage. Based on the number and status of connected playback AR devices and the total number of trials to be performed, it assigns playbacks to each connected device. Visual and inertial input data are transferred to local storage on each of those AR devices via an Android Debug Bridge `push' command. Finally, the module calculates a \textit{wait time} based on when 
playback will complete (known because sequences must be run in real time).

\noindent \textbf{Sequence playback: }The sequence playback module replays the virtual and inertial input data on the playback AR device(s) the number of times specified by the sequence assignment module, and outputs a trajectory estimate for each iteration. We implemented this module in an AR app, created in Unity 2021.3.14f1, using AR Foundation 4.2. The app is installed on each playback AR device that will be used to replay sequences. Throughout sequence playback the estimated device pose is logged every frame, and the final trajectory estimates are saved to local storage after each iteration.

\noindent \textbf{Uncertainty-based error estimation: }The uncertainty-based error estimation module calculates error estimates for each sub-trajectory from the trajectory estimates generated by the sequence playback module(s). It is implemented on the server, within the same Python app as the sequence scheduling module. After using the \textit{wait time} provided by the sequence assignment module to wait for sequence playback to complete, this module transfers trajectory estimates from the playback AR device(s) to the server via an Android Debug Bridge `pull' command. It then calculates pose error estimates for each sub-trajectory using the method described in Section~\ref{subsec:Uncertainty-basedEstimationMethod}, with trajectory evaluations performed using the evo Python package \cite{grupp2017evo}. Finally, the set of sub-trajectory error estimates (with associated timestamps) is transmitted to the user AR device via HTTP. 


\noindent \textbf{Trajectory visualization: }
Once a trajectory is created, the trajectory visualization module renders the initial trajectory estimate. The user AR device positions recorded at the endpoints and midpoints of each sub-trajectory define the endpoints of cylinders and the position of connecting spheres (diameters of 0.3m) showing the trajectory path. The AR device orientations recorded at each midpoint are used to render camera frustums at the sub-trajectory midpoints immediately following 0.5m increments on the trajectory -- this indicates which direction the camera was facing at this point in the trajectory (a possible future enhancement would be to take into account rotation over the entire sub-trajectory when rendering frustums). Once sub-trajectory error estimates are obtained from the uncertainty-based error estimation module, they are loaded into a dictionary. The trajectory visualization module queries this dictionary to fetch estimated error for each timestamped sub-trajectory, and coloring or annotations can be applied to each rendered sub-trajectory based on these error values. Depth estimates from the trajectory creation module can also be used to inform the rendering of other types of visualization, as described in Section~\ref{subsec:VisualizationTechniques}.

\subsection{System Latency Characterization}
\label{subsec:SystemLatencyCharacterization}
We identify three potential latency bottlenecks for SiTAR, depending on the system configuration. First, the most obvious latency source is the \textit{sequence playback module}, due to the need to replay sequences in real time. Sequence playback latency $L_p$ can be approximated by $L_p = d \times \lceil T / D \rceil$, 
where $d$ is the duration of the trajectory (sequence) created by the user, $T$ is the number of trials, 
and $D$ is the number of playback AR devices. Thus we can greatly reduce the playback latency by increasing the number of playback AR devices
, but at the minimum it will be the trajectory duration.

Second, the data transmission step that incurs the greatest latency is the \textit{wireless transfer of visual and inertial input data from the user AR device to the server}. In our testing the average file size of a sequence recorded with the ARCore Recording and Playback API is approximately 1--1.2MB per second of recorded data; assuming typical transmission speeds, transmission latency to either an edge or cloud server will usually be negligible compared to playback latency. For example, in our in-the-wild study (Section~\ref{sec:VisualizationsforSituatedTrajectoryAnalysis}) the mean latency of transferring sequences to an edge server was 2.9s. 


Third, we consider the latency of the \textit{uncertainty-based error estimation module} on the server, largely determined by the number of trials used for estimation and the length of estimated trajectories. Recalling from Section~\ref{subsec:Uncertainty-basedEstimationMethod} that for $n$ trials we perform $n(n-1)/2$ trajectory evaluations, latency grows substantially with both the number of trials and trajectory length, because error must be calculated for each 
sub-trajectory in each evaluation. We measure mean latency (10 trials) for a long sequence, \mbox{TUM VI} room1, a medium-length sequence, SenseTime A3, and our own short sequence recorded using ARCore, when running trajectory evaluations on a desktop computer (Intel i7-9700K CPU, Nvidia GeForce RTX 2060 GPU); our results are shown in Table~\ref{tab:Latency}. We note the modest estimation performance gains achieved with e.g., 50 rather than 20 trials (Figure~\ref{fig:UncertaintyEstimationResults}) may not justify the 
latency overhead in many scenarios.

We also measure the \textit{end-to-end latency of SiTAR} when the backend is deployed using an edge server, with the hardware shown in Figure~\ref{fig:SystemImage}. These measurements are from 45 trials we conducted as part of an in-the-wild study (Section~\ref{sec:VisualizationsforSituatedTrajectoryAnalysis}), across various residential and organizational local area networks; trajectory duration was approximately 15s, five trials were used for pose error estimation, the uncertainty calculation setting was a trimmed mean at 20\%, and we used one playback AR device. Mean end-to-end latency was 89.1s, with a standard deviation of 12.9s and a range of 60.3--132.7s.

\begin{table}[t]
  \centering
  \caption{Mean latency 
  of the 
  uncertainty-based estimation module. 
  }
  \vspace{-0.12in}
  \begin{tabular}{|P{2.94cm}|P{0.38cm}|P{0.50cm}|P{0.50cm}|P{0.65cm}|P{0.8cm}|}
  \hline
\multirow{2}{2.94cm}{\centering \textbf{Sequence (duration)}} & \multicolumn{5}{c|}{\textbf{\# of trials used for estimation}} \\
\cline{2-6}
 & \textbf{5} & \textbf{10} & \textbf{20} & \textbf{50} & \textbf{100}\\
\hline
\textbf{TUM VI room1 (141s)}&4.5s&18.9s&78.9s&539.2s&2809.3s\\
\hline
\textbf{SenseTime A3 (45s)}&1.9s&8.0s&32.7s&229.4s&1275.2s\\
\hline
\textbf{Ours (15s)}&0.5s&1.9s&8.1s&67.7s&525.4s\\
\hline
\end{tabular}
  \label{tab:Latency}
  \vspace{-0.6cm}
\end{table}

\section{Visualizations for Situated Trajectory Analysis}
\label{sec:VisualizationsforSituatedTrajectoryAnalysis}
In this section we present three possible visualization techniques for situated trajectory analysis, and evaluate them through an in-the-wild user study. Rather than comparing user preferences in a fully controlled lab study, our goal here is to investigate the challenges presented by the diverse environments in which situated trajectory analysis will be conducted, and thereby inform future techniques. 

\subsection{Visualization Techniques}
\label{subsec:VisualizationTechniques}
First, we detail the three visualization techniques we created and implemented. These visualizations apply different types of semantic enrichment to device trajectories to convey estimated pose error magnitude and problematic environment regions to the user.\hfill \break

\noindent \textbf{Trajectory-only: }Here we apply pose error-based semantic enrichment through coloring of the originally rendered trajectory. Once the pose error values are obtained for each sub-trajectory, we apply different colors to the cylinders and spheres used to render that sub-trajectory. For example, for the three-class classifier we developed in Section~\ref{subsec:Uncertainty-basedEstimationEvaluation}, which categorizes sub-trajectories into high, medium and low estimated error, we render the sub-trajectories associated with high error in red, medium error in yellow, and low error in the original blue. Any camera frustums rendered within a sub-trajectory are also rendered in the appropriate color. Examples of our trajectory-only visualization are shown in Figure~\ref{fig:VisualizationA}.

\noindent \textbf{Trajectory + exclamation points: }This technique extends our trajectory-only technique by rendering 3D exclamation points, a symbol widely associated with a warning \cite{edworthy1996warning}, in problematic environment regions. To this end, we calculate the point on an environment surface that the user AR device camera is facing in the middle of each sub-trajectory. We experimented with ray casting against 
the environment map generated by the VI-SLAM algorithm; however, on devices without a 
depth sensor there are often large gaps in this map, particularly in the problematic regions (e.g., blank walls) we wish to detect \cite{sartipi2020deep}. Instead, we leverage the depth map from the ARCore Depth API to obtain the distance from the camera to the environment. Combined with the device pose, this gives us the 3D position on the 
surface where we render a virtual object 
\cite{du2020depthlab}. We check if an exclamation point was previously rendered within a Euclidean distance of 0.5m (to avoid clusters of exclamation points
), and if not, render a 3D exclamation point (width=5cm, height=30cm, depth=5cm)
, coloring it to match the sub-trajectory. Examples of our trajectory + exclamation points technique are shown in Figure~\ref{fig:VisualizationB}.

\noindent \textbf{Trajectory + warning signs: }This technique also extends our \mbox{trajectory-only} technique, but instead of rendering exclamation points on problematic environment surfaces, we render a triangular 2D warning sign containing an exclamation point, a widely known symbol which when employed in heads-up displays for driver assistance was associated with low driver response times \cite{cheng2007active}. Not only do we render the warning signs at the position defined by the environment surface, but we also match the orientation of the sign to that 
surface, so it appears, for example, pasted onto a wall or table. We again leverage the depth map generated by the ARCore Depth API, but this time we obtain both the depth magnitude and normal estimate, which combined with the camera pose provides the required 6D pose. As with exclamation points, we check that a warning sign was not previously rendered within a Euclidean distance of 0.5m, then render a 2D warning sign (width=30cm, height=30cm), and color it to match the sub-trajectory. Examples of our trajectory + warning signs technique are shown in Figure~\ref{fig:VisualizationC}.\hfill \break

\subsection{Preliminary Testing}
\label{subsec:PreliminaryTesting}
After integrating our visualization techniques into the SiTAR trajectory visualization module  (Section~\ref{subsec:SystemDesign}), we performed approximately 50 trials in five environments, including lab, office and residential spaces. We qualitatively assessed 
performance by comparing the 
problematic 
regions SiTAR detected with expert judgments of visual input data likely to result in high pose error (based on both qualitative statements in previous works, e.g., \cite{jinyu2019survey, campos2021orb}, and our 
experience). For example, we expected the system to identify featureless regions such as blank walls and powered-off TVs, poorly-lit areas, and reflective surfaces such as glass; as shown in Figure~\ref{fig:Teaser}, these are the types of regions SiTAR highlighted. This consistency of performance with our earlier testing (Section~\ref{subsec:Uncertainty-basedEstimationEvaluation}) is logical, as to the best of our knowledge the VI-SLAM 
algorithms employed by commercial AR platforms share many similarities 
with the open-source algorithm on which we developed our estimation method.

Importantly, this testing indicated that in the vast majority of cases pose error magnitude is not so great as to render visualizations 
away from the intended problematic regions, such that the guidance they provide is still reliable. This was supported by the data from 
our in-the-wild study (Section~\ref{subsec:VisualizationExperimentsDesign}); the average sub-trajectory error for the 45 trajectories ranged from 0.5cm to 24.9cm, with a mean of 3.7cm. These results emphasize the presence of noticeable errors, but the magnitudes observed mean that visualizations should still be rendered in the same environment region. To warn users when error is high enough to potentially affect the reliability of situated visualizations, we added additional UI guidance on the SiTAR user AR device when average estimated pose error exceeds 20cm.

Based on our testing, we also made two small adjustments to the SiTAR output to support our user study. 
First, to control task length we limit the number of problematic environment regions identified to three. Second, to maximize the probability that at least three regions are found (and avoid having to repeat trials), we search sub-trajectory pose error estimates for magnitudes greater than or equal to 10cm, 5cm, 2cm, then 1cm, until three sub-trajectories are found.


\subsection{Visualization Experiments Design}
\label{subsec:VisualizationExperimentsDesign}
To evaluate our three visualizations, we conducted an IRB-approved user study. As our goal was to investigate challenges associated with deploying different visualization techniques for situated trajectory analysis in diverse environments, we opted for an in-the-wild study, in which participants used SiTAR in a variety of different locations. We chose a within-subject design in which each participant experienced all three visualizations, presented in randomized order.\hfill \break

\noindent \textbf{Participants and locations: }We recruited 15 participants (9 female, 6 male; aged 18 to 44; normal or corrected-to-normal eyesight) from our personal and professional networks. Our study covers 13 diverse environments, including a wildlife center, a medical simulation space, a library, a botanical garden, a museum, a classroom, plus multiple residential and office spaces. This included outdoor spaces and indoor spaces ranging 
from small offices ($\approx$ 3m$\times$4m$\times$3m) to large open lobbies ($\approx$ 20m$\times$8m$\times$6m). Examples of these environments are shown in Figure~\ref{fig:Teaser}. One environment was used per participant, so that each participant experienced the three visualizations under similar conditions, but different areas and movement paths were used for each visualization to avoid learning effects. 

\noindent \textbf{Task: }Given that our targeted use case for situated trajectory analysis is the identification of problematic environment regions, we define a task in which participants have to locate and capture images of three problematic environment regions conveyed by each visualization. This task is similar to the search task used in studies on locating out-of-view objects in AR (e.g., \cite{schinke2010visualization,gruenefeld2018beyond,wieland2022arrow}). Prior to the task, users created a 15-second trajectory (timing guided by the study administrator) that included a view of multiple environment regions, then found a position from which they could view the whole trajectory. As soon as the error-based visualization was rendered, participants were notified with UI text and an audio effect that the task had started. During the task the participants were free to locate the three problematic environment regions identified in any order, and were required to press a UI button to capture an image of each region. Once they had captured the three images participants were notified with UI text and an audio effect that the task had been completed.

\noindent \textbf{Dependent variables and operationalization: }We measured task performance using task completion time, the total amount of time taken to capture an image of each of the three problematic environment regions. We measured user experience using the User Experience Questionnaire (UEQ) \cite{laugwitz2008construction}, along with subjective rankings obtained from a post-experiment questionnaire. In this post-experiment questionnaire we also captured participants reasoning for these subjective rankings by asking them to justify their choices with advantages and disadvantages of the visualizations. We measured the workload associated with each visualization using the raw, unweighted  NASA Task Load Index (NASA-TLX) \cite{hart1988development}.

\begin{figure}[b]
\centering
\captionsetup[subfigure]
{justification=centering}
\vspace{-0.3cm}
\begin{subfigure}{.22\textwidth}
  \includegraphics[width=1.0\linewidth]{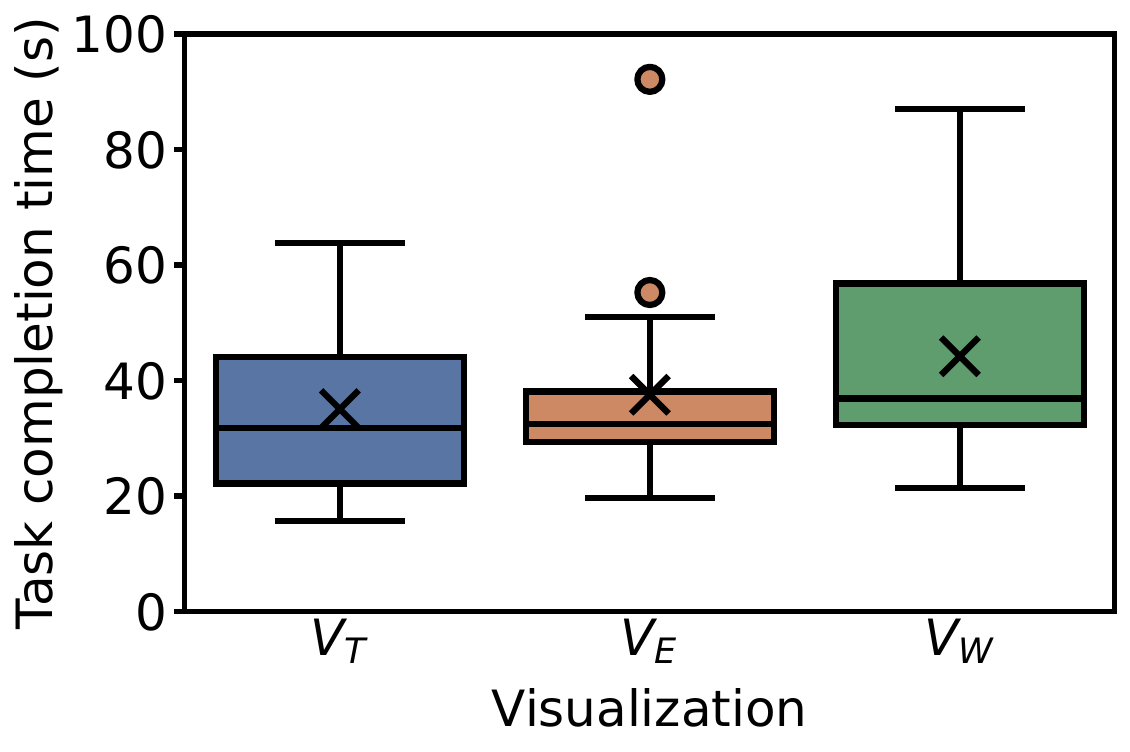}
  \vspace{-0.6cm}
  \caption{}
  \label{fig:Taskcompletiontime}
\end{subfigure}
\hspace{0.5cm}%
\begin{subfigure}{.205\textwidth} \includegraphics[width=1.0\linewidth]{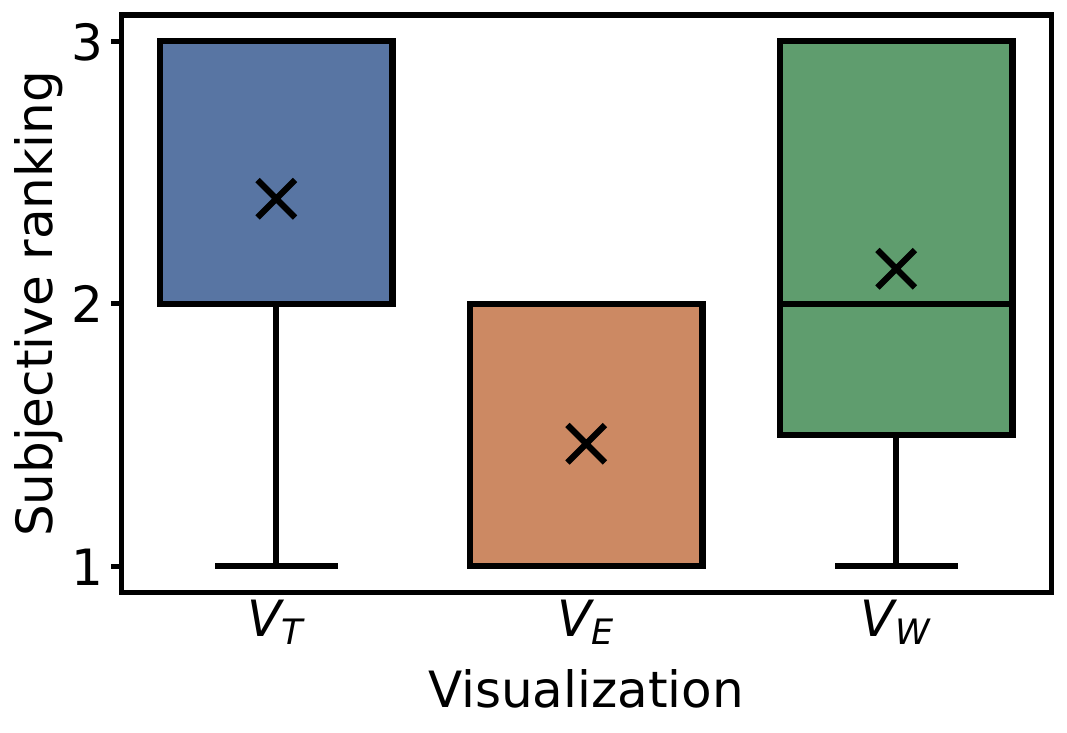}
  \vspace{-0.6cm}
  \caption{}
  \label{fig:Subjectiveranking}
\end{subfigure}
\vspace{-0.4cm}
\caption{Results from our in-the-wild study of three visualizations for situated trajectory analysis: (a) performance, measured using task completion time; (b) subjective ranking (lower is better).}
\label{fig:TimeSubjective}
\end{figure}

\noindent \textbf{Apparatus: }
For the user AR device in SiTAR we used a Google Pixel 7 Pro running Android 13 and ARCore v1.37, with the display set at full brightness; participants used this device to create trajectories, perform tasks, and complete questionnaires. We deployed our SiTAR system backend (the uncertainty-based pose error estimation) on an edge server, an Apple MacBook Pro laptop with a 2.6GHz dual-core Intel Core i5 processor and 8GB RAM, running macOS BigSur 11.7 and Python 3.8. We used one playback AR device, a Google Pixel 7 running Android 13 and ARCore v1.37.

\noindent \textbf{Procedure: }
After introducing participants to the study we asked them to sign a consent form. They then completed a pre-experiment questionnaire, used to obtain data on demographics and AR experience. Next, we conducted a training phase, in which we introduced participants to the app 
and the three visualizations, provided a demonstration of trajectory creation and task completion, and allowed the participant to practice this until they were comfortable completing it independently. We then started the main study, in which participants conducted one trial 
with each visualization. After each trial, participants completed the UEQ and NASA-TLX for the visualization they had just experienced. At the end of the study the participants completed the post-experiment questionnaire.

\begin{figure*}[t]
\begin{subfigure}{1.05\columnwidth}
\includegraphics[width=\columnwidth]{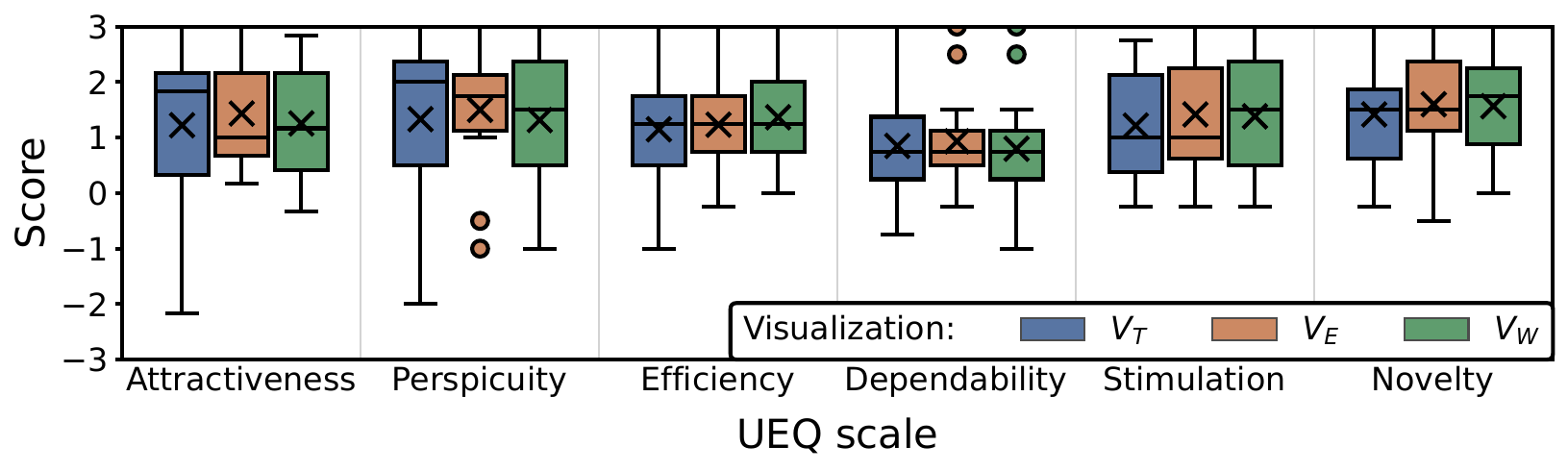}%
\vspace{-0.3cm}
\caption{User Experience Questionnaire (UEQ)}%
\label{fig:UEQ}%
\end{subfigure}
\begin{subfigure}{1.05\columnwidth}
\includegraphics[width=\columnwidth]{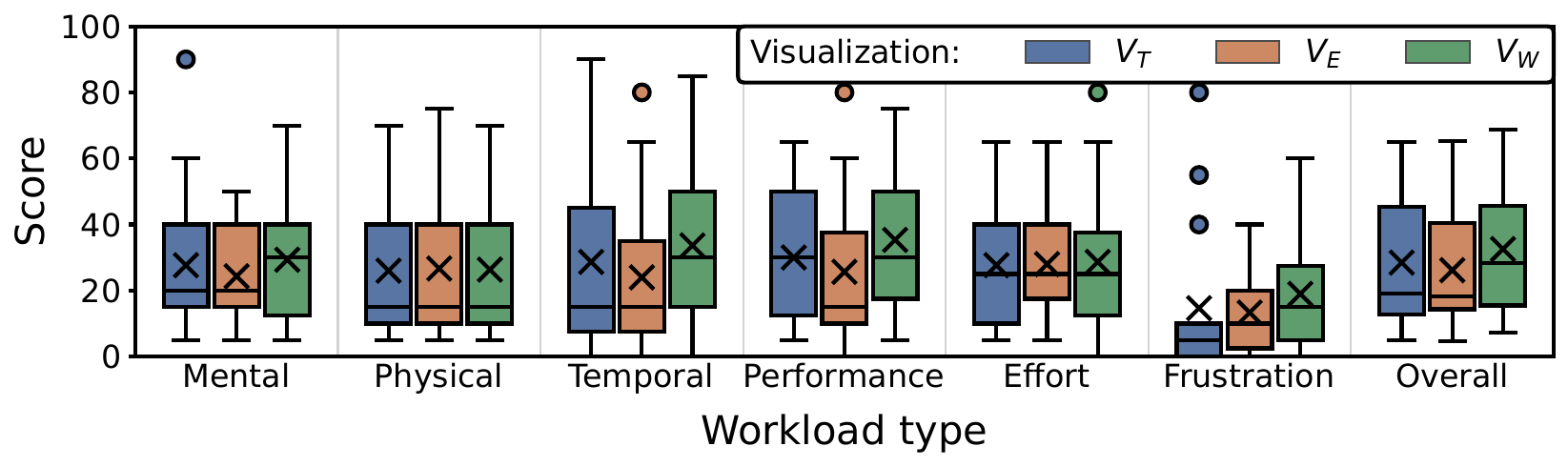}%
\vspace{-0.3cm}
\caption{NASA Task Load Index (NASA-TLX)}%
\label{fig:TLX}%
\end{subfigure}\hfill%
\vspace{-0.7cm}
\caption{Results from our in-the-wild study of three visualizations for situated trajectory analysis: (a) user experience, measured using the user experience questionnaire \cite{laugwitz2008construction}; (b) workload, measured using the NASA Task Load Index \cite{hart1988development}.}
\label{UserStudy}
\vspace{-0.5cm}
\end{figure*}

\subsection{Visualization Experiments Results}
\label{subsec:VisualizationExperimentsResults}
The results for our three visualization techniques are shown in Figures~6 and 7. Below we compare them for each of our dependent variables -- task performance, user experience, and workload. For readability, we denote the trajectory-only visualization by $V_T$, trajectory + exclamation points by $V_E$, and trajectory + warning signs by $V_W$. To test for statistical significance we use Friedman's ANOVA (suitable for our repeated measures design with three conditions, without assuming data is normally distributed), followed by post-hoc pairwise comparisons using the Nemenyi test. For statistical significance we assume $\alpha = 0.05$. Our comparisons should be interpreted in the context that this is an exploratory rather than a fully controlled study; the different environments in which participants experienced visualizations could affect them differently, and while we found no evidence of order effects, they are possible in this randomized rather than counterbalanced study. However, the following results provide insights into the effectiveness of our visualizations across diverse scenarios, informing the design of future visualizations and studies.

\subsubsection{Task Performance}
\label{subsubsec:TaskPerformance}
For task completion time (Figure~\ref{fig:Taskcompletiontime}), Friedman's ANOVA revealed significant differences between conditions ($\altmathcal{X}^2(2)=6.533, p = 0.038$), and a post-hoc Nemenyi test showed task completion time was significantly greater for $V_W$ than $V_T$ ($p = 0.046$, $V_W$ mean $= 44.2$s, $V_T$ mean $= 35.1$s). Our post-experiment survey revealed one possible reason, in that some participants indicated they found the warning signs hard to locate, one commenting that
``\textit{the warning signs could be harder to determine where they were}.'' This 
may be due to the warning signs being 2D, and hence not visible from certain directions. Other participants noted that even when visible, warning signs did not convey the desired information effectively: 
``\textit{the orientation/location of the warning signs were confusing and it was not clear which signs aligned with which parts of the trajectory}.''

Upon examining instances of high task completion times for $V_T$ and $V_W$, we found they occurred either in large environments or regions with glass surfaces (e.g., windows, museum display cases). For example, in the case of the two outliers for $V_E$, at least one exclamation point was rendered $>6$m away. These types of environments pose significant challenges for depth estimate-based rendering of virtual objects
, because estimates are less accurate at greater distances and on reflective surfaces \cite{zhang2022indepth}. This in turn results in the inaccurate position or orientation of virtual objects; for example, an exclamation point may be rendered `outside' of a glass window,  
or a warning sign at 90$^{\circ}$ to the actual surface of a museum display case. This degradation of 
visualizations motivates the development of techniques more robust to large environments and reflective surfaces, especially given their prevalence in modern architecture.


\subsubsection{User Experience}
\label{subsubsec:UserExperience}
For subjective rankings (Figure~\ref{fig:Subjectiveranking}), Friedman's ANOVA revealed significant differences between conditions ($\altmathcal{X}^2(2)=6.933, p = 0.031$), and a post-hoc Nemenyi test showed users ranked $V_E$ significantly higher (lower subjective ranking score) than $V_T$ ($p = 0.029$, $V_E$ mean $= 1.5$, $V_T$ mean $= 2.4$). Participant justifications 
included ``\textit{I liked the additional signal that indicated a problem area}'' and ``\textit{Really hard to gauge without...symbols if you were pointing the right direction}'', indicating that they liked the extra information that the exclamation points and warning signs provided. Participants who ranked warning signs lower than exclamation points often noted the 
problem of locating the signs or determining their orientation.

Friedman's ANOVA did not reveal significant differences between conditions for any of the UEQ scales (Figure~\ref{fig:UEQ}), perhaps due to all visualizations having a prominent element, the trajectory, in common. 
We note that while all three visualizations obtained mean scores $>1$ on the attractiveness, perspicuity, efficiency, stimulation and novelty scales, 
participants rated them lower for dependability, with mean scores for all visualizations $<1$. This indicates that there is work to be done on improving the perceived reliability of situated trajectory analysis, and ensuring users feel in control of interactions.

\subsubsection{Workload}
\label{subsubsec:Workload}
For the NASA TLX results (Figure~\ref{fig:TLX}), Friedman's ANOVA revealed significant differences between conditions for overall workload ($\altmathcal{X}^2(2)=6.241, p = 0.044$), and a post-hoc Nemenyi test showed overall workload was significantly higher for $V_W$ than $V_E$ ($p = 0.046$, $V_W$ mean $= 32.6$, $V_E$ mean $= 26.2$). This is consistent with participant feedback on difficulty finding warning signs or that their orientation was confusing. While three participants scored $V_T$ highly for frustration -- comments included ``\textit{Just the trajectory did not provide much information}'' and ``\textit{Trajectory has insufficient data to help me understand my task}'' -- all other participants scored $V_T \leqslant 10$ for frustration, with the median frustration score for $V_T$ lower than for $V_E$ or $V_W$. This indicates that despite participants preferring visualizations which attached virtual objects to problematic regions, searching for those objects may be a source of frustration, and prompts the inclusion of techniques that make them more noticeable (e.g., the addition of motion in kineticons \cite{harrison2011kineticons}), and to help users navigate to them when they are out of view (e.g., \cite{schinke2010visualization,gruenefeld2018beyond,wieland2022arrow}).


\section{Conclusion and Future Work}
\label{sec:ConclusionsandFutureWork}
In this paper we presented SiTAR, 
the first system that incorporates pose error estimates into situated trajectory analysis. To enable this, we developed the first VI-SLAM pose error estimation method based on uncertainty propagation, and demonstrated its efficacy on four VI-SLAM datasets. We integrated our pose error estimation method into our SiTAR system, which can be deployed on real AR devices with an edge or cloud server backend. We tested SiTAR in an in-the-wild study with 15 users in 13 diverse environments; this study revealed an overall preference for pose error visualizations which placed virtual objects in problematic environment regions, as well as how environment properties can impact these techniques.  

SiTAR opens up a number of opportunities for future work. 
First, our visualization experiments in diverse environments prompt investigation into how techniques can be best adapted to suit different types of environments; this includes the visualizations themselves, sub-trajectory length, and the distance between virtual objects attached to environment regions. Second, there are a number of possible improvements to situated pose error visualizations; these include techniques to aid users further in determining the source of pose error (e.g., visualizations that show the entire spatial regions covered by input camera images, the addition of input data properties to trajectories), visualizations more robust to user pose, and methods of indicating which environment regions have been scanned. Backup methods of communicating problematic regions (when tracking error is so great as to make situated visualizations unreliable) should also be considered, e.g., displaying the input camera frames for a sub-trajectory. Finally, there is exciting scope for extending SiTAR to analyze trajectories from multiple users. 
For example, we envision recording user trajectories, evaluating them with the SiTAR backend offline, then generating persistent situated visualizations that combine error estimates from multiple trajectories. Administrators can then analyze these data at their convenience, to identify environment or content adjustments that will improve virtual object stability.

\acknowledgments{
We thank our study participants for their 
assistance in this research. This work was supported in part by NSF grants CSR-1903136, CNS-1908051 and CNS-2112562, NSF CAREER Award IIS-2046072, a Meta Research Award and a CISCO Research Award.}

\bibliographystyle{abbrv-doi}

\bibliography{references}
\end{document}